\documentclass[a4paper]{article}

\usepackage{graphicx}
\usepackage{amssymb}
\usepackage{amsmath}
\usepackage{cite}
\usepackage[left=2cm,top=2cm,right=2cm,bottom=2cm,nohead]{geometry}
\usepackage[pdftex,unicode,pdfstartview=FitH,colorlinks=true,pdfborder={0 0 0 0},urlcolor=blue]{hyperref}
\usepackage[labelsep=space,format=plain,margin=0.5cm,textformat=simple,justification=justified,small,parskip=0cm]{caption}
\usepackage[onehalfspacing]{setspace}

\title{Consentaneous agent-based and stochastic model of the financial markets}
\author{Vygintas Gontis\thanks{\href{mailto:vygintas@gontis.eu}{vygintas@gontis.eu}; \url{http://gontis.eu/en/}}, Aleksejus Kononovicius\thanks{\href{mailto:aleksejus.kononovicius@gmail.com}{aleksejus.kononovicius@gmail.com}; \url{http://kononovicius.lt/en/}}}
\date{}

\newcommand{\rmd}{\mathrm{d}}

\begin{document}

\maketitle

\begin{abstract}
We are looking for the agent-based treatment of the financial markets considering necessity to build bridges between microscopic, agent based, and macroscopic, phenomenological modeling. The acknowledgment that agent-based modeling framework, which may provide qualitative and quantitative understanding of the financial markets, is very ambiguous emphasizes the exceptional value of well defined analytically tractable agent systems. Herding as one of the behavior peculiarities considered in the behavioral finance is the main property of the agent interactions we deal with in this contribution. Looking for the consentaneous agent-based and macroscopic approach we combine two origins of the noise: exogenous one, related to the information flow, and endogenous one, arising form the complex stochastic dynamics of agents. As a result we propose a three state agent-based herding model of the financial markets. From this agent-based model we derive a set of stochastic differential equations, which describes underlying macroscopic dynamics of agent population and log price in the financial markets. The obtained solution is then subjected to the exogenous noise, which shapes instantaneous return fluctuations. We test both Gaussian and q-Gaussian noise as a source of the short term fluctuations. The resulting model of the return in the financial markets with the same set of parameters reproduces empirical probability and spectral densities of absolute return observed in New York, Warsaw and NASDAQ OMX Vilnius Stock Exchanges. Our result confirms the prevalent idea in behavioral finance that herding interactions may be dominant over agent rationality and contribute towards bubble formation.
\end{abstract}

\section{Introduction}

Statistical physics has got the edge over socio-economic sciences in the understanding of complex systems \cite{Aoki2007Cambridge,Ball2012Springer,Pietronero2008EPN,Stauffer2011Corr,Tsallis2009Springer,Castellano2009RevModPhys,Scalas2009Economics}. This happened due to the fact that physicists were able to start from the understanding of simple phenomena via simple models and later built the complexity up together with the increasing complexity of the considered phenomena. On the other hand socio-economic sciences had to face complexity right from the start as socio-economic systems are in no way simple systems - they are intrinsically complex at many different levels at the same time. Financial markets are one of the most interesting examples of such complex systems. Unlike in physics we have no direct way to gain insights into the nature of microscopic interactions in financial markets, thus the understanding of the financial market fluctuations may become rather limited and very ambiguous. Yet the understanding might be improved indirectly through the further development of the complex systems approach \cite{Farmer2012EPJ}. First of all, currently there are huge amounts of the available empirical data, which itself is attracting representatives of the experimental sciences \cite{Mantegna2000Cambridge}. Also there is an agent-based modeling framework, which may provide qualitative and quantitative understanding of the financial markets. The intense applications of these ideas is still ongoing \cite{Mantegna2000Cambridge,Voit2005Springer,Roehner2010SciCul} and the challenge is still open.

Agent-based modeling has become one of the key tools, which could improve the
understanding of the financial markets as well as lead to the potential applications \cite{Bouchaud2008Nature, Farmer2009Nature, Lux2009NaturePhys,Schinckus2013ConPhys,Cristelli2012Fermi,Chakraborti2011RQUF2}.
Currently there are many differing agent-based approaches in the modeling of the financial markets.
Some of them aim to be as realistic as possible, yet they usually end up being
too complex to posses analytical treatment. One of the most prominent examples of these kind of models is
the Lux-Marchesi model \cite{Lux1999Nature}. The more recent approaches in the similar
direction consider modeling order books \cite{Lye2012PhysA,Preis2006EPL,Schmitt2012EPL}. Other
approaches, on the other hand, aim to capture the most general properties of the many complex
socio-economic systems (some of the examples include \cite{Gekle2005EPJB,Traulsen2012PhysRevE}).
Though there are also some interesting approaches which combine realism and analytical tractability, e.g.
Feng \emph{et al.} \cite{Feng2012PNAS} have used both empirical data and trader survey data to
construct agent-based and stochastic model for the financial market.
Looking for the ideal agent-based approach we would consider as a primary necessity
to build bridges between microscopic, agent based, and macroscopic, phenomenological, modeling\cite{Kononovicius2012IntSys}. Following this trace of thought it would be rational to combine two origins of the noise: exogenous one, related to the information flow, and endogenous one, arising form the complex stochastic dynamics of agents.
Such integral view of the financial markets can be achieved only with very simple zero-intelligence
agent-based models and macroscopic, phenomenological, approaches incorporating external information flow.
This is the main idea of our present consideration of the financial markets.

The expected properties of such model lead us to the return fluctuations, characterized by the
power law distributions and the power law autocorrelations of absolute return considered in \cite{Lux1996AFE, Cont1997Springer,Gopikrishnan1998EPJB,
Ding1993JEF,Vandewalle1997PhysA,Lobato1998JBES,Lux2002Springer}. We investigate an agent-based herding model of the financial markets, which proves to be
rather realistic and also simple enough to be analytically tractable \cite{Kononovicius2012PhysA,
Kononovicius2013EPL}. Namely we consider a three agent states' model \cite{Kononovicius2013EPL}
and incorporate it into the standard model of the stock price described by the geometric Brownian motion or into process with statistical feedback \cite{Borland1998PhysRevE}, exhibiting Tsallis statistics. We find that
the improved three state agent-based herding model reproduces the power law statistics
observed in the empirical data extracted from the NYSE Trades and Quotes database, Warsaw Stock Exchange and NASDAQ OMX Vilnius Stock Exchange.

We start by discussing the possible alternatives in macroscopic and phenomenological modeling providing some insight into the possible connection to the agent-based microscopic approach. Next we develop the microscopic approach by defining the herding interactions between three agent groups and incorporate it into a consistent model of the financial markets. Further we couple the endogenous fluctuations of the agent system with the exogenous information flow noise incorporated in macroscopic approach and provide detailed comparison with the empirical data. Finally we discuss the obtained results in the context of the proposed double stochastic model of the return in the financial markets.

\section{Methods}

\subsection{Macroscopic and phenomenological versus microscopic and agent-based treatment of the financial markets}
It is the natural peculiarity of the social systems to be treated first of all from the macroscopic and phenomenological point of view.
In contrast to the natural sciences microscopic treatment of the social systems is ambiguous and hardly can be considered as a starting point for the consistent modeling.
The complexity of human behavior leaves us without any opportunity to consider human agent in action as a determined dynamic trajectory.  The financial markets as an example of
the social behavior first of all are considered as a macroscopic system exhibiting stochastic movement of the variables such as asset price, trading volume or return \cite{Voit2005Springer}.
Despite lack of knowledge regarding microscopic background of the financial systems there is considerable progress in stochastic modeling producing very practical applications \cite{Jeanblanc2009Springer,Taylor1994MFinance}.
The standard model of stock prices, $S(t)$, referred to as geometric Brownian process, is widely accepted in financial analysis
\begin{equation}
\rmd S = \mu(t) S \rmd t + \sigma(t) S \rmd W. \label{eq:GBprice}
\end{equation}
In the above Wiener process $W$ can be considered as an external information flow noise while $\sigma(t)$ accounts for the stochastic volatility. Though one must consider the model of stock prices following geometric Brownian motion as a hypothesis which has to be checked critically, this serves as a background for many empirical studies and further econometric financial market model developments. Acknowledgment that analysis taking $\mu(t)$ and $\sigma(t)$ constant have a finite horizon
of application has become an important motivation for the study of the ARCH and GARCH processes \cite{Campbell1997Princeton,Engle1982Econometrica,Bollerslev1986Econometrics} as well as for the stochastic modeling of volatility $\sigma(t)$ \cite{Taylor1994MFinance}.

We acknowledge this phenomenological approach as a good starting point for the macroscopic financial market description incorporating external information flow noise $W$ and we will go further by modeling volatility $\sigma(t)$ as an outcome of some agent-based herding model. The main purpose of this approach is to demonstrate how sophisticated statistical features of the financial markets can be reproduced by combining endogenous and exogenous stochasticity.

Here we consider only the most simple case, when $\sigma(t)$ fluctuations are slow in comparison with external noise $W$. In such case the return, $r_t (T)=\ln \frac{S(t+T)}{S(t)}$, in the time period $T$ can be written as a solution of Eq. (\ref{eq:GBprice})
\begin{equation}
r_t (T) = \left(\mu(t)-\frac{1}{2} \sigma(t)^2 \right) T + \sigma(t) W(T).\label{eq:GBreturn}
\end{equation}
This equation defines instantaneous return fluctuations as a Gaussian random variable with mean $(\mu-\frac{1}{2} \sigma^2) T$ and variance $\sigma^2 T$. Let us exclude here from the
consideration long term price movements defined by the mean $(\mu-\frac{1}{2} \sigma^2) T$ as we will define the dynamics of price from microscopic agent-based part of model. This assumption means that we take from phenomenological model only the general idea how to combine exogenous and  endogenous noise. Then Eq. (\ref{eq:GBreturn}) simplifies
to the instantaneous Gaussian fluctuations $ N[0,\sigma(t)^2 T] $  with zero mean and variance $ \sigma(t)^2 T $.

In \cite{Gontis2010Intech,Gontis2010PhysA}, while relying on the empirical analysis, we have assumed that the return, $r_t (T)$, fluctuates as instantaneous q-Gaussian noise $N_q[r_0(x),\lambda]$ with some power-law exponent $\lambda = \frac{2}{q-1} = 5$, and driven by some stochastic process $x(t)$ defining second parameter of fluctuations $r_0(x)$. $r_0(x)$ was introduced as a linear function of absolute return moving average $|x|$ calculated from some nonlinear stochastic model \cite{Gontis2010PhysA}
\begin{equation}
r_0(x) = b + a |x|, \label{eq:r0fun}
\end{equation}
where parameter $b$ serves as a time scale of exogenous noise and $\frac{b}{a}$ quantifies the relative input of exogenous noise in comparison with endogenous one described by $|x(t)|$.

A more solid background for this kind of approach can be found in the work by L. Borland \cite{Borland2002PRL}. The idea to replace geometric Brownian process of market price by process with statistical feedback \cite{Borland1998PhysRevE} leads to the equation of return $r_t (T)$ as function of time interval $T$ given by
\begin{equation}
\rmd r_t(T) =  \sigma(t) \rmd \Omega_T,\label{eq:QGreturn}
\end{equation}
where $\Omega_T$ evolves according to the statistical feedback process \cite{Borland1998PhysRevE}
\begin{equation}
\rmd \Omega_T =  P(\Omega,T)^{\frac{(1-q)}{2}} \rmd W_T,\label{eq:Omegaeq}
\end{equation}
and $P(\Omega,T)$ satisfies the nonlinear Fokker-Planck equation
\begin{equation}
\frac{\partial}{\partial T} P(\Omega,T) =  \frac{\partial}{\partial \Omega^2} P^{2-q}(\Omega,T).\label{eq:OmegaFP}
\end{equation}
The explicit solution for $P$ in the region of $q$ values $1<q<5/3$ can be written as one of the Tsallis distributions \cite{Tsallis1996PhysRevE}
\begin{equation}
P_q(\Omega,T) = \frac{1}{C(T)} \left\{1-\gamma(T)(1-q)\Omega^2\right\}^{\frac{1}{1-q}}, \label{eq:TsallisD}
\end{equation}
where $C(T)$ and $\gamma(T)$ are as follows
\begin{eqnarray}
& C(T) = \left[\frac{\sqrt{\pi} \Gamma(\frac{3-q}{2(q-1)})}{\sqrt{q-1} \Gamma(\frac{1}{q-1})}\right]^{\frac{2}{3-q}} \left[(2-q)(3-q)T \right]^{\frac{1}{3-q}}, \nonumber \\
& \gamma(T) = \left[\frac{ \Gamma(\frac{1}{q-1})}{\sqrt{\pi} \Gamma(\frac{3-q}{2(q-1)})}\right]^{\frac{2(q-1)}{3-q}} \frac{(q-1)^{\frac{q-1}{3-q}}}{\left[(2-q)(3-q)T \right]^{\frac{2}{3-q}}}. \label{eq:Cgamma}
\end{eqnarray}
Assuming $\sigma(t)$ as slow stochastic process in comparison with $\Omega_T$ from Eq. (\ref{eq:QGreturn}) one gets that $r_t(T)=\sigma(t) \Omega_T$. This sets PDF for $r_t(T)$ the same as for $\Omega_T$ Eq. (\ref{eq:TsallisD}), one just has to replace $\Omega$ and, $C(T)$ defined in Eqs. (\ref{eq:Cgamma}) by $\frac{r_t(T)}{\sigma(t)}$ and $\sigma(t) C(T)$ accordingly.
This gives a Tsallis distribution for $r_t(T)$ as
\begin{equation}
P_q(r_t(T)) = \frac{1}{\sigma(t) C(T)} \left\{1-(1-q)\frac{r_t(T)^2}{(3-q)\sigma(t)^2 \sigma_q(T)^2}\right\}^{\frac{1}{1-q}}, \label{eq:Tsallisr}
\end{equation}
where $\sigma_q(T)$ as new parameter related to previous one $\gamma(T)$ can be written as
\begin{eqnarray}
& \sigma_q(T) = \left[\sqrt{\pi\frac{3-q}{q-}}\frac{ \Gamma(\frac{3-q}{2(q-1)})}{ \Gamma(\frac{1}{q-1})}\right]^{\frac{(q-1)}{3-q}} \left[(2-q)T \right]^{\frac{1}{3-q}}=\\
& =\left[\sqrt{\pi (\lambda-1)}\frac{ \Gamma(\frac{\lambda-1}{2})}{ \Gamma(\frac{\lambda}{2})}\right]^{\frac{1}{\lambda-1}} \left[\frac{(\lambda-2)}{\lambda}T \right]^{\frac{\lambda}{2(\lambda-1)}}. \label{eq:sigmaq}
\end{eqnarray}

 Now we are prepared to combine two phenomenological approaches introduced by Eqs. (\ref{eq:GBreturn}) and (\ref{eq:QGreturn}) with agent-based endogenous three state herding model. $\sigma(t)$ serves as a measure of system volatility in both of the phenomenological approaches. It is reasonable to assume that financial market is in the lowest level of possible volatility when assets market value $P(t)$ is equal to the it's fundamental value $P_f(t)$, lets define it as constant $\sigma(t) = b$. Volatility of financial system increases when market value of the asset deviates from  the fundamental value. These deviations  can be accounted as $p(t) = \ln \frac{P(t)}{P_f(t)}$. Further in this contribution we will assume that volatility $\sigma(t)$ is defined by $|p(t)|$ through the linear relation
\begin{equation}
\sigma(t) = b(1 + a |p(t)|), \label{eq:sigmaprice}
\end{equation}
where parameter $b$ serves as a scale of exogenous noise and $a$ quantifies the relative input of endogenous noise. Both parameters $a$ and $b$ have to be defined from empirical data. To complete the model we have to propose agent-based consideration of log price $p(t)$. In the following section we present the three state herding model giving stochastic equations for the log price $p(t)$.

\subsection{The three state herding model as a source for the endogenous stochastic dynamics}

Having discussed a macroscopic view of the financial fluctuations we now switch to the microscopic consideration of the endogenous fluctuations.
Let us derive the system of stochastic differential equations defining the endogenous log-price fluctuations from a setup of appropriate agent groups composition.
We consider a system of N heterogenous agents - market traders continually changing their trading strategies between three possible choices: fundamentalists,
chartists optimists and chartists pessimists. We further develop this commonly used agent group setup \cite{Feng2012PNAS} by considering all transitions
between three agent states to be a result of binary herding interactions between agents during their market transactions.

Fundamentalists are traders with fundamental understanding of the true stock value, which is commonly quantified as the stock's fundamental price, $P_f(t)$.
We exclude from our consideration any movements of the fundamental price. The assumption of constant
fundamental price means that we further will consider price fluctuations around its fundamental value. Taking into account
a long-term rational expectations of the fundamentalists their excess demand, $ED_f(t)$, might be assumed to be given by \cite{Alfarano2005CompEco}
\begin{equation}
ED_f(t) = N_f(t) \ln \frac{P_f}{P(t)} ,
\end{equation}
where $N_f(t)$ is a number of the fundamentalists inside the market and $P(t)$ is a current market price. The rationality of fundamental traders keeps
asset price around its fundamental value as they sell when $P_f < P(t)$, and buy when $P_f > P(t)$.

Short term speculations cause unpredictable price movements. There is a huge variation of speculative trading strategies, so it is rational, from the statistical physics point of view, to consider these variations as statistically irrelevant.
We make only two distinctions between chartists: optimists suggest to buy and pessimist suggest to sell at a given moment. Thus the excess demand of the chartist traders,
$ED_c(t)$, can be written as:
\begin{equation}
ED_c(t) = \bar{r} [N_o(t) - N_p(t)] ,
\end{equation}
where $\bar{r}$ is a relative impact factor of the chartist trader and further will be integrated into a certain empirical parameter, $N_o$ and $N_p$ are the total numbers of optimists and pessimists respectively. So, as you can see, we replace a big variety of "rational" chartist trading strategies by herding kinetics between just two options: buy or sell.

The proposed system of heterogenous agents defines asset price movement by applying the Walrasian scenario. A fair price reflects the current supply and demand and the Walrasian scenario in its contemporary form may be written as
\begin{equation}
\frac{1}{\beta N} \frac{\rmd p(t)}{\rmd t} = - n_f(t) p(t) + \bar{r} [n_o(t) - n_p(t)] ,
\end{equation}
here $\beta$ is a speed of the price adjustment, $N$ a total number of traders in the market, $p(t) = \ln \frac{P(t)}{P_f}$ and $n_i(t) = \frac{N_i(t)}{N}$. By assuming that the number of traders in the market is large, $N \rightarrow \infty$, one obtains:
\begin{equation}
p(t) = \bar{r} \frac{n_o(t) - n_p(t)}{n_f(t)} .\label{eq:logprice}
\end{equation}

Stochastic dynamics of the proposed agent-based system is defined by the occupations of the three agent states:
\begin{equation}
n_f = \frac{N_f}{N}, \quad n_p = \frac{N_p}{N}, \quad n_o = \frac{N_0}{N}.
\end{equation}

\begin{figure}
\centering
\includegraphics[width=7cm]{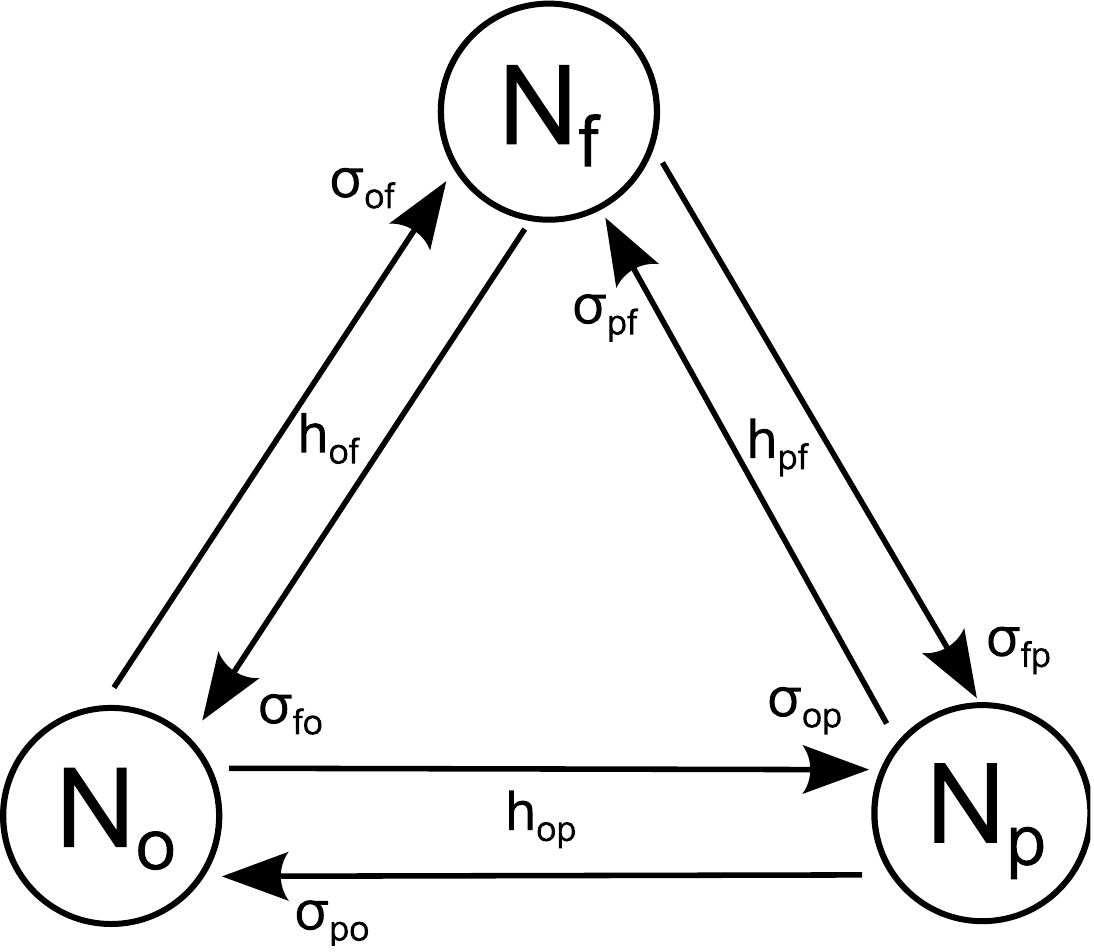}
    \caption{{Schematic representation of the three state herding model, where relevant parameters are shown.} The arrows point in the directions of the possible transitions, each of the transitions pairs is modeled using original Kirman's model.}
    \label{figure1}
\end{figure}

One can model the evolution of occupations as a Markov chain with some reasonable assumptions for the sake of simplicity. There are six one agent transition possibilities in three group setup, see Fig. \ref{figure1} and \cite{Kononovicius2013EPL}.  Few assumptions are natural for the financial markets as there is some symmetry. With the notation of agent transitions from state $i$ to $j$ as subscripts to any parameter $A_{ij}$, where i and j take values from the set $\{f,c,o,p\}=$ $\{\textmd{fundamentalists, all chartists, optimists, pessimists}\}$, we will use following assumptions, for the rates of spontaneous transitions: $\sigma_{op}=\sigma_{po}=\sigma_{cc}$, $\sigma_{fp}=\sigma_{fo}=\sigma_{fc}/2$, $\sigma_{pf}=\sigma_{of}=\sigma_{cf}$, and for the herding transitions: $h_{fp}=h_{pf}=h_{fo}=h_{of}=h$. Finally it is reasonable to assume that transitions between chartist states are much faster than between chartists and fundamentalists $h_{op}=h_{po}=H h$, $H \gg 1$, $\sigma_{cc} \gg \sigma_{cf}$, $\sigma_{cc} \gg \sigma_{fc}$. Taking into account the restraint $N_f+N_c=N$ and having in mind that transitions $f \rightarrow o$ are equivalent to $f \rightarrow p$, one can  can write one step herding transition rates between $f$ and $c$ groups for given $n_f=N_f/N$ and $n_c=N_c/N$  as \cite{Kononovicius2012PhysA}
\begin{eqnarray}
& \pi_{fc}(n_f) = n_f \left[\frac{\sigma_{fc}}{N}+h (1-n_f)\right] , \nonumber \\
& \pi_{cf}(n_f) = (1-n_f) \left[\frac{\sigma_{cf}}{N}+h n_f\right] . \label{eq:kirmratefc}
\end{eqnarray}
Here transition rates have the same form as in Kirman's herding model \cite{Kirman1993QJE}. Using Eq. (\ref{eq:kirmratefc}) one can write the Master equation for PDF $P(n_f,t)$ and derive Fokker-Planck equation in the limit $N \rightarrow \infty$ \cite{Kononovicius2012PhysA}. One dimensional Fokker-Planck equations has its equivalent stochastic differential equation which for the $n_f$ can be written as
\begin{equation}
\rmd n_f = \left[ (1-n_f) \sigma_{cf} - n_f \sigma_{fc} \right] \rmd t + \sqrt{2 h n_f (1-n_f)} \rmd W_f.  \label{eq:sdenf}
\end{equation}
The next step in this approach is to define dynamics of $n_p$ under restraint $n_f + n_p + n_o = 1$. An adiabatic approximation assuming variable $n_f$ changes slowly in comparison with $n_p$ or $n_o$ is helpful here. This enables to consider $n_p$ dynamics as one dimensional process as well. Let us write the transition rates for $n_p$ in the same way as in Eq. (\ref{eq:kirmratefc})
\begin{eqnarray}
& \pi_{po}(n_p) = n_p \left[\frac{\sigma_{cc}}{N}+ H h (1-n_f-n_p)\right] , \nonumber \\
& \pi_{op}(n_p) = (1-n_f-n_p) \left[\frac{\sigma_{cc}}{N}+ H h n_p\right] . \label{eq:kirmratecc}
\end{eqnarray}
As in other similar cases \cite{Kononovicius2012PhysA} these one step transitions lead to the SDE for $n_p$
\begin{equation}
\rmd n_p = (1-n_f - 2 n_p) \sigma_{cc} \rmd t + \sqrt{2 H h n_p (1-n_f-n_p)} \rmd W_p.  \label{eq:sdenp}
\end{equation}
Equations (\ref{eq:sdenf}) and (\ref{eq:sdenp}) form a system of coupled SDEs and define the agent population dynamics in three state agent model taking into account the previous assumptions. It is possible to rewrite Eq. (\ref{eq:sdenp}) in the form without direct dependance on $n_f$ by introducing another variable $\xi(t) = \frac{n_o(t)-n_p(t)}{n_o(t)+n_p(t)}$, average mood of the chartists, instead of $n_p$. Such variable substitution makes second SDE independent of the first one
\begin{equation}
\rmd \xi = - 2 \sigma_{cc} \xi \rmd t + \sqrt{2 H h(1-\xi^2)} \rmd W_{\xi} . \label{eq:sdexi}
\end{equation}
Equations (\ref{eq:sdenf}) and  (\ref{eq:sdexi}) are independent and define occupation dynamics of the same three state agent-based herding model.

In the previous work we considered generalization of the herding model introducing variable interevent time $\tau$, see \cite{Kononovicius2012PhysA,Kononovicius2013EPL}. It is a natural feedback of a macroscopic state on the microscopic behavior, activity of agents. Such feedback is an empirically defined phenomena and can be quantified through the relation of trading volume with return \cite{Rak2013APP}. We introduce this feedback into proposed three state herding model as a trading activity, rate of transactions, $\frac{1}{\tau(n_f,\xi)}$ defined as
\begin{equation}
\frac{1}{\tau(n_f,\xi)}= \left( 1 + a \left| \frac{1-n_f}{n_f} \xi \right| \right)^{\alpha} =\left( 1 + a \left| p(t) \right| \right)^{\alpha}, \label{eq:taunfxi}
\end{equation}
where empirical parameter $a$ is the same as in Eq. (\ref{eq:sigmaprice}).
Such power-law behavior is consistent with empirical data quantifying relation of short term return $r$ with trading volume $V$, $V(r) \sim r^\alpha$, where $\alpha \simeq 2$, \cite{Gabaix2003Nature,Gabaix2006QJE,Farmer2004QF}. Notice that for the geometric Brownian motion model of stock price the return as increment of log-price is proportional to the log-price. This is the another reason together with search of simplicity we use log-price instead of return in Eq. (\ref{eq:taunfxi}).

Taking into account the discussed feedback mechanism, introducing scaled time, $t_s=h t$, and appropriately redefining model parameters: $\varepsilon_{cf} = \sigma_{cf} / h$, $\varepsilon_{fc} = \sigma_{fc} / h$, $\varepsilon_{cc} = \sigma_{cc} / (H h)$,  we are able to rewrite stochastic differential equations of endogenous model as
\begin{eqnarray}
& \rmd n_f = \left[ \frac{(1-n_f) \varepsilon_{cf}}{\tau(n_f,\xi)} - n_f \varepsilon_{fc} \right] \rmd t_s + \sqrt{\frac{2 n_f (1-n_f)}{\tau(n_f,\xi)}} \rmd W_{s,f} , \label{eq:nftau}\\
& \rmd \xi = - \frac{2 H \varepsilon_{cc} \xi}{\tau(n_f,\xi)} \rmd t_s + \sqrt{\frac{2 H (1-\xi^2)}{\tau(n_f,\xi)}} \rmd W_{s,\xi} , \label{eq:xitau}
\end{eqnarray}
Then log-price defined in Eq. (\ref{eq:logprice}) can be expressed through the stochastic variables of the model $n_f$ and $\xi$
\begin{equation}
p(t) = \frac{1 - n_f(t)}{n_f(t)} \xi(t) \label{eq:logpricenfxi}.
\end{equation}
In Eq. (\ref{eq:logpricenfxi}) and further we omit parameter $\bar{r}$ as consider it integrated into parameter $a$.

This concludes the definition of consentaneous agent-based and stochastic model of the financial markets as Eq. (\ref{eq:logpricenfxi}) defines joint endogenous and exogenous volatility $\sigma(t)$ introduced by Eq. (\ref{eq:sigmaprice}).

\subsection{Numerical algorithm}
We solve Eqs. (\ref{eq:nftau}) and (\ref{eq:xitau}) by using Euler-Maruyama method \cite{Kloeden1999Springer} with variable time step, $\Delta t_i$. Namely we have transformed a set of stochastic differential equations into a set of difference equations:
\begin{eqnarray}
& x_{i+1} = x_i + h \left[ \frac{(1-x_i) \varepsilon_{cf}}{\tau(x_i,\xi_i)} - x_i \varepsilon_{fc} \right] \Delta t_i + \sqrt{\frac{2 h x_i (1-x_i)}{\tau(x_i,\xi_i)} \Delta t_i} \zeta_{1,i} , \label{eq:difnf} \\
& \xi_{i+1} = \xi_i - \frac{2 h H \varepsilon_{cc} \xi_i}{\tau(x_i,\xi_i)} \Delta t_i + \sqrt{\frac{2 h H (1- \xi_i^2)}{\tau(x_i,\xi_i)} \Delta t_i} \zeta_{2,i} , \label{eq:difxi} \\
& t_{i+1} = t_i + \Delta t_i , \quad \Delta t_i = \frac{\kappa^2 \tau(x_i,\xi_i)}{h (1 + \varepsilon_{cf} + \varepsilon_{fc} + H (1+ 2 \varepsilon_{cc}))} .
\end{eqnarray}
In the above we have changed the notation, $x_i = n_{f,i}$, to improve readability of the difference equations. While $\zeta_{j,i}$ stands for the uncorrelated normally distributed random variables with zero mean and unit variance. Also note that in the difference equations above we have introduced additional parameter $\kappa$, which is responsible for the precision of the numerical results. The smaller $\kappa$ value gets, the more precise numerical simulations are, but the longer computation time grows. During the numerical simulations we found $\kappa = 0.03$ to be the optimal value precision-wise and time-wise.

In order to keep $n_f$ and $\xi$ well defined we introduce absorbing boundaries near the edges of the intervals in which these variables are defined. Namely, we require that each $n_f$ belongs to $[\delta , 1 - \delta]$ and $\xi$ belongs to $[-1+ \delta, 1- \delta]$. In other words before using $x_{i+1}$ and $\xi_{i_1}$ obtained from the Eqs. (\ref{eq:difnf}) and (\ref{eq:difxi}), we apply $\mathrm{min}(\dots)$ and $\mathrm{max(\dots)}$ functions on them:
\begin{eqnarray}
& x'_{i+1} = \mathrm{min}(\mathrm{max}(x_{i+1},\delta),1-\delta) , \\
& \xi'_{i+1} = \mathrm{min}(\mathrm{max}(\xi_{i+1},-1+\delta),1-\delta) .
\end{eqnarray}
The new $x'_{i+1}$ and $\xi'_{i+1}$ are certainly well defined and may be used in further simulations. In our simulations we have used $\delta = 10^{-6}$.

Next step in our numerical simulation is to obtain time series of $p_i$, discretized at one minute time periods, and apply $q$-Gaussian,
\begin{equation}
r(t,T) = b \left( 1 + a \left|\frac{1-x_t}{x_t} \xi_t \right| \right) \sigma_q(T) N_q \left(0, 1\right) ,
\end{equation}
or, in certain cases, Gaussian,
\begin{equation}
r(t,T) = b \left( 1 + a \left|\frac{1-x_t}{x_t} \xi_t \right| \right) \sqrt{T} N \left(0, 1\right) ,
\end{equation}
noise on it. Afterwards we normalize the obtained one minute return time series so in the end it would have unit variance.

Finally we add as many one minute returns as we need to obtain the final return time series at a desired discretization intervals. In the second section we analyze empirical data from New York, Warsaw and NASDAQ OMX Vilnius Stock Exchanges and reproduce it's statistical properties in the numerical simulation.

\section{Results and Discussion}

\subsection{Comparison of model simulation and empirical data}

Now we will adjust model parameters to reproduce empirical data of the return in three different markets. The model return $r_t(T)$ in the time interval $T$ can be written as
\begin{equation}
r(t,T) = \sigma(t) \sqrt{T} N(0,1), \label{eq:Gaussreturn}
\end{equation}
for the Gaussian external noise, see Eq. (\ref{eq:GBreturn}),  and
\begin{equation}
r(t,T) = N_q(0,\sigma(t) \sigma_q(T)), \label{eq:qGaussreturn}
\end{equation}
for the q-Gaussian one, see Eq.(\ref{eq:Tsallisr}). Here $N(0,1)$ denotes normally distributed random variable with zero mean and unit variance, and $N_q(0,\sigma(t) \sigma_q(T))$ denotes Tsallis random variable distributed as defined by Eq.(\ref{eq:Tsallisr}). We choose $T=1 \quad \textmd{minute}$ as primary sufficiently short time interval where $\sigma(t)$ fluctuations are negligible and solve equations (\ref{eq:nftau}) and (\ref{eq:xitau}) numerically in successive time intervals to get 1 minute time series of $p(t)$, Eq. (\ref{eq:logpricenfxi}). From the price time series one can produce time series for the volatilities $\sigma(t)$ and returns $r(t,T)$ using Eq. (\ref{eq:Gaussreturn}) or Eq. (\ref{eq:qGaussreturn}).

We compare model return series with empirical return time series extracted from high frequency trading data on New York, Warsaw, and NASDAQ OMX Vilnius Stock Exchanges. These series were transformed into successive sequences of empirical 1 minute returns. Produced empirical return series were normalized by standard deviation calculated on the entire time sequence of selected stock. For this comparison with empirical data we select only a few stocks from each stock exchange, which have more or less constant long term average trading activity to avoid considerable input of possible trend into time series statistics. From NYSE data we have selected stocks: BMY, GM, MO, T, traded for 27 months from January, 2005. From Warsaw SE stocks: TPSA, KGHM, traded from Novemver 2000 to January 2014, and PZU traded from May 2010 to January 2014. From NASDAQ OMX Vilnius data stocks: APG1L, IVL1L, PTR1L, SAB1L, TEO1L, traded from May 2005 to December 2013. We do not extend this analysis into more wide representation of stocks as only few stocks from NASDAQ OMX Vilnius are liquid  enough for such analysis. Comparable demonstration of  general features across markets is the main purpose of this consideration. In every stock group series of different stocks are considered as separate realizations of the same stochastic process and so we present empirical statistical information as average over realizations (stocks).

\begin{figure}
\centering
\includegraphics[width=15cm]{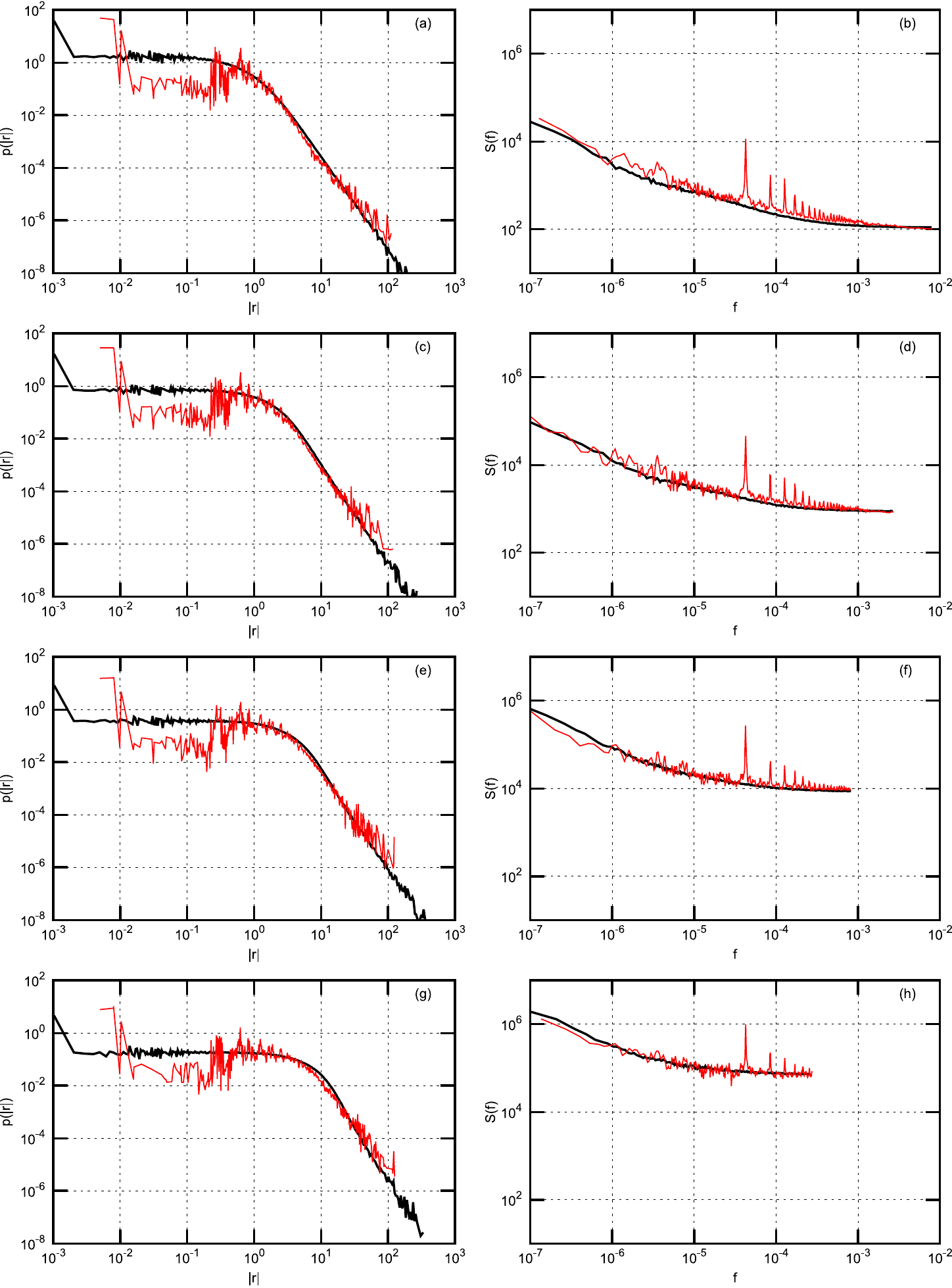}
    \caption{{Probability density function (PDF) and power spectral density (PSD) of return with $q$-Gaussian noise for NYSE stocks: BMY, GM, MO, T.}  Empirical (red) and model (black) PDF (first column) and PSD (second column). (a) and (b) - 1 minute; (c) and (d) - 3 minutes; (e) and (f) - 10 minutes; (g) and (h) - 30 minutes. Model parameters are as follows: $\varepsilon_{cf}=0.1$, $\varepsilon_{fc}=3$, $\varepsilon_{cc}=3$, $H=300$, $h=10^{-8} s^{-1}$, $a=0.5$, $b=1$, $\lambda=4$, $\alpha=2$.}
    \label{figure2}
\end{figure}

\begin{figure}
\centering
\includegraphics[width=15cm]{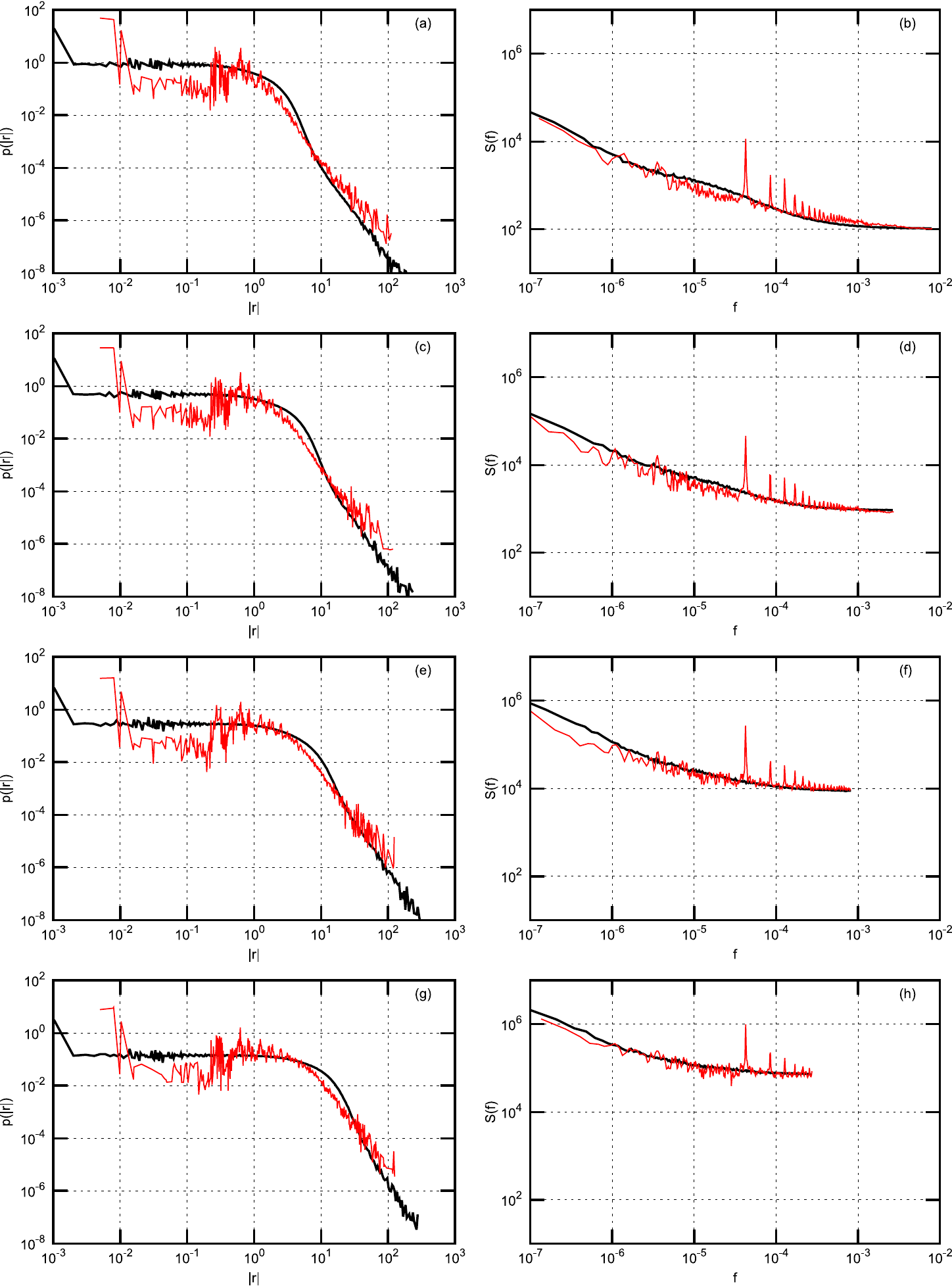}
    \caption{{Probability density function (PDF) and power spectral density (PSD) of return with Gaussian noise for NYSE stocks: BMY, GM, MO, T.}  Empirical (red) and model (black) PDF (first column) and PSD (second column). (a) and (b) - 1 minute; (c) and (d) - 3 minutes; (e) and (f) - 10 minutes; (g) and (h) - 30 minutes. Model parameters are as follows: $\varepsilon_{cf}=0.1$, $\varepsilon_{fc}=3$, $\varepsilon_{cc}=3$, $H=300$, $h=10^{-8} s^{-1}$, $a=0.5$, $b=1$, $\alpha=2$.}
    \label{figure3}
\end{figure}

\begin{figure}
\centering
\includegraphics[width=15cm]{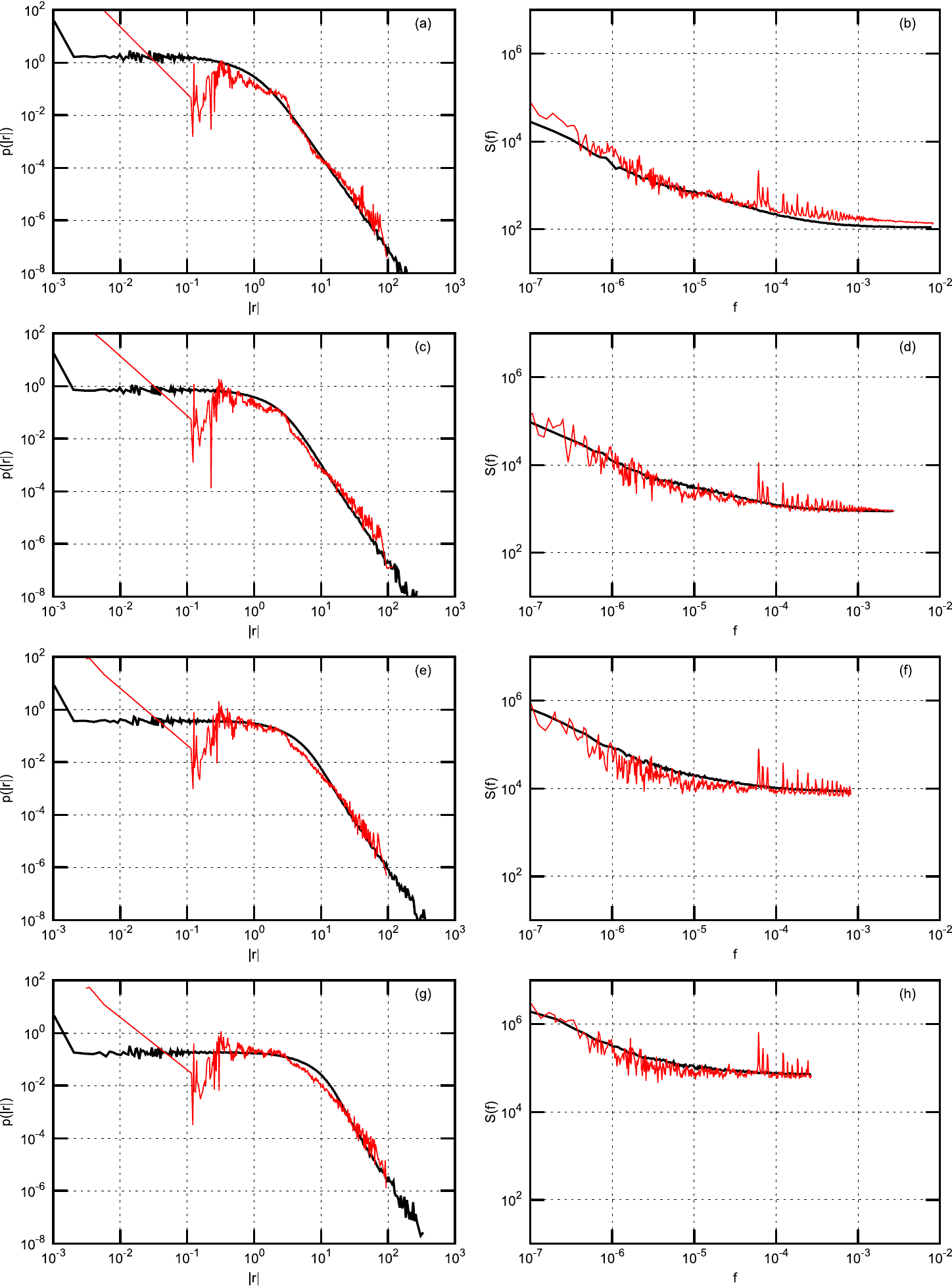}
    \caption{{Probability density function (PDF) and power spectral density (PSD) of return with $q$-Gaussian noise for Warsaw Stock Exchange  stocks: KGHM, PZU, TPSA.}  Empirical (red) and model (black) PDF (firs column) and PSD (second column). (a) and (b) - 1 minute; (c) and (d) - 3 minutes; (e) and (f) - 10 minutes; (g) and (h) - 30 minutes. Model parameters are the same as in Fig. \ref{figure2}.}
    \label{figure4}
\end{figure}

\begin{figure}
\centering
\includegraphics[width=15cm]{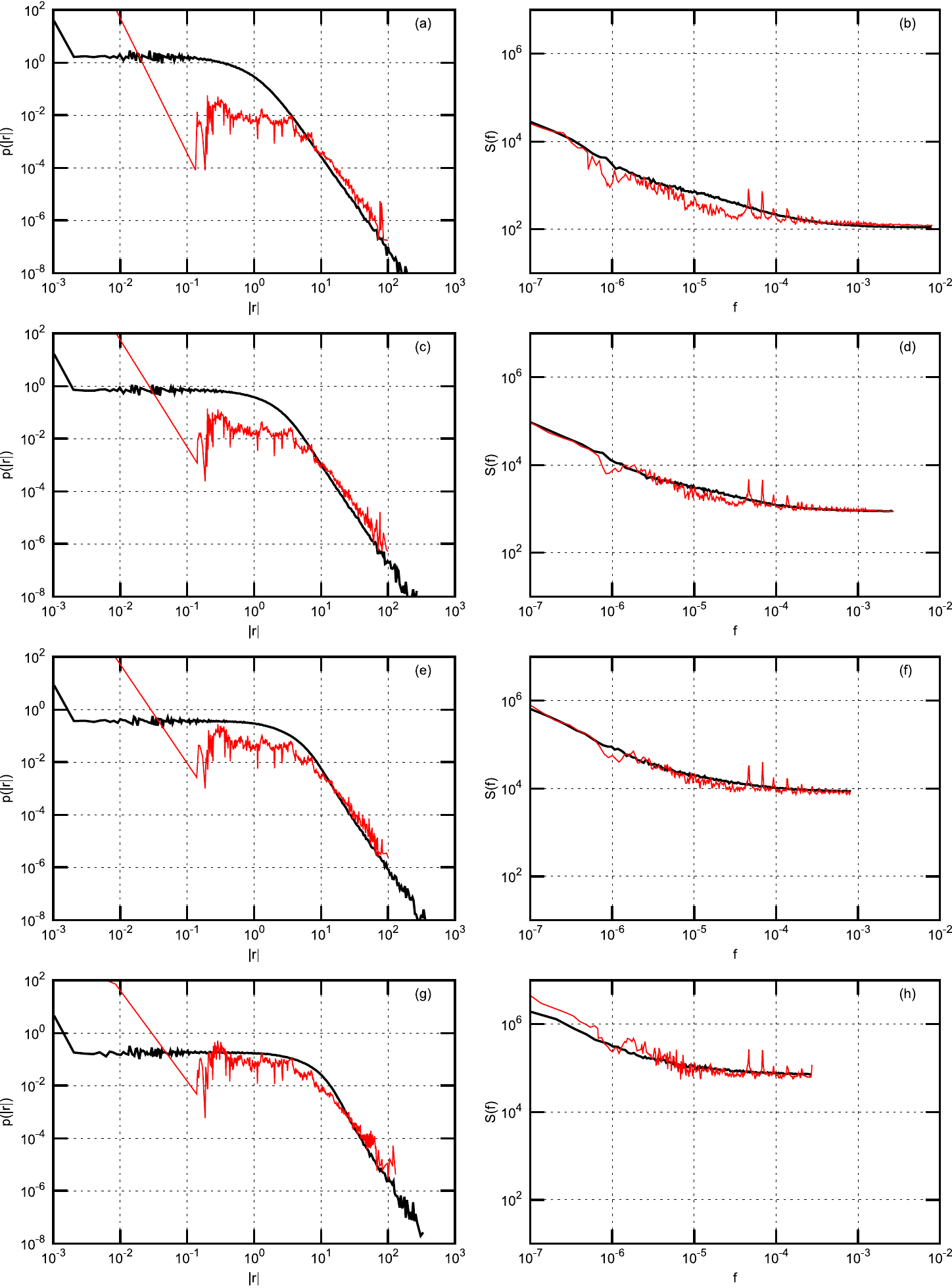}
    \caption{{Probability density function (PDF) and power spectral density (PSD) of return with $q$-Gaussian noise for NASDAQ OMX Vilnius stocks:  APG1L, IVL1L, PTR1L, SAB1L, TEO1L.}  Empirical (red) and model (black) PDF (first column) and PSD (second column). (a) and (b) - 1 minute; (c) and (d) - 3 minutes; (e) and (f) - 10 minutes; (g) and (h) - 30 minutes. Model parameters are the same as in Fig. \ref{figure2}.}
    \label{figure5}
\end{figure}

In order to compare scaling of the statistics with increasing time interval $T$ of return definition we just sum 1 minute returns of model and empirical time series in successive  intervals of 3 minutes, of 10 minutes or 30 minutes. For each stock exchange considered and each time interval $T$ we calculate absolute return probability density function (PDF) and power spectral density (PSD). Both PDF and PSD are obtained by averaging over stocks in the considered stock exchanges. Results are presented in four figures: \ref{figure2} - for NYSE data; \ref{figure3} - for NYSE data with Gaussian noise; \ref{figure4} - for Warsaw Stock Exchange data; \ref{figure5} - for NASDAQ OMX Vilnius data. In figures: \ref{figure2}, \ref{figure4}, \ref{figure5} empirical data (red lines) are compared with model statistics (black line) calculated with the same choice of parameters: $\varepsilon_{cf}=0.1$, $\varepsilon_{fc}=3$, $\varepsilon_{cc}=3$, $H=300$, $h=10^{-8} s^{-1}$, $a=0.5$, $b=1$, $\lambda=4$, $\alpha=2$.

From our point of view the achieved coincidence of empirical and model statistics is better than expected. PDF and PSD coincide almost for all markets and all time $T$ scales. q-Gaussian noise suites much better for this model, compare Fig. \ref{figure2} calculated with q-Gaussian noise and Fig. \ref{figure3} calculated with Gaussian one. The use of q-Gaussian noise can be confirmed by more detailed study of empirical data as well. 

\begin{figure}
\centering
\includegraphics[width=15cm]{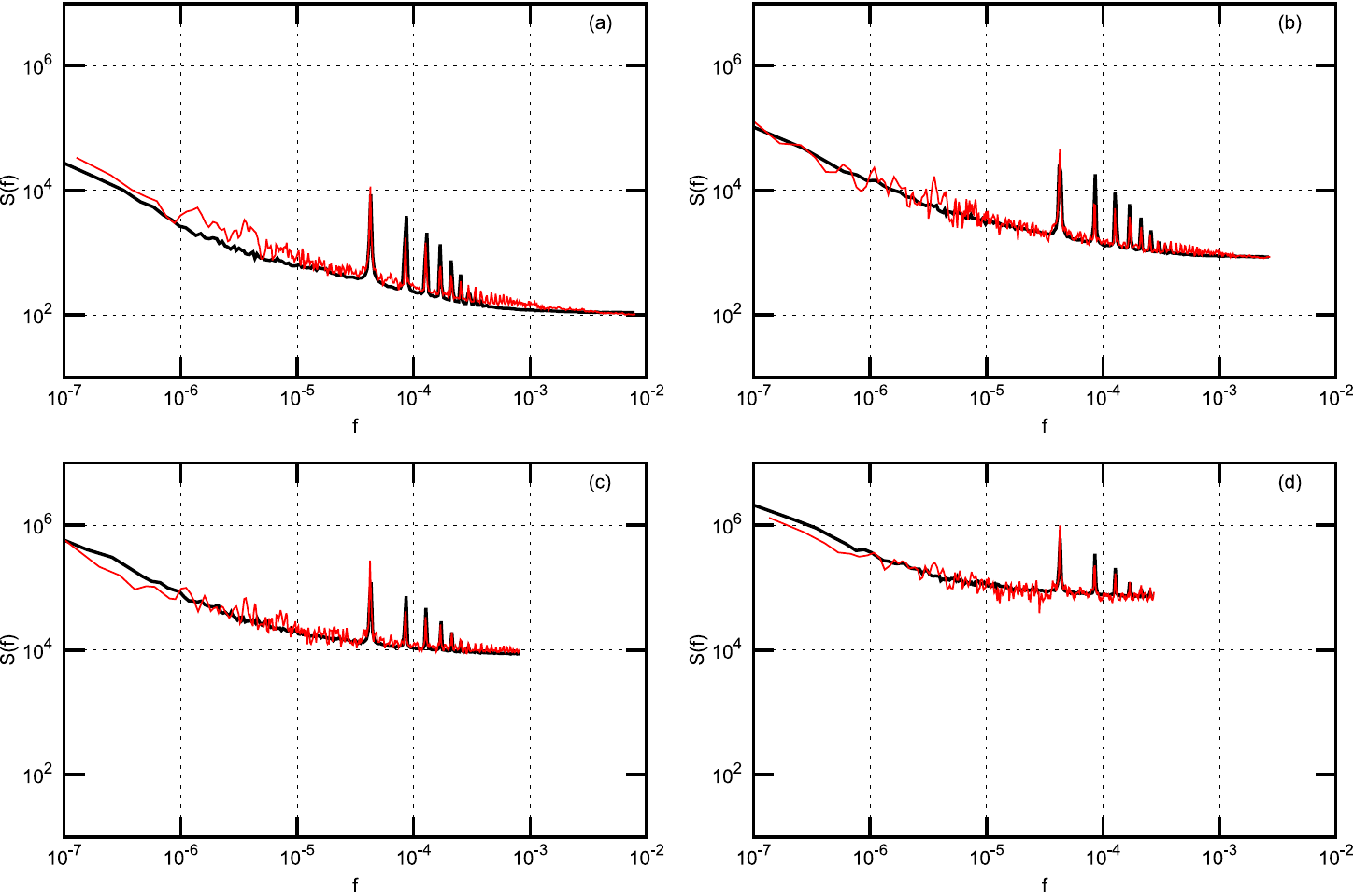}
    \caption{{Power spectral density (PSD) of return with $q$-Gaussian noise and seasonality accounted for NYSE stocks:  BMY, GM, MO, T.}  Empirical (red) and model (black) PSD accounting for parameter's $b$ seasonality Eq. (\ref{eq:bseasonality}) . (a) - 1 minute; (b) - 3 minutes; (c) - 10 minutes; (d) - 30 minutes. Other model parameters are the same as in Fig. \ref{figure2}.}
    \label{figure6}
\end{figure}

There are some observed discrepancies of empirical and model results, which have reasonable explanations. For example, spikes observed in empirical PSD are related with one trading day seasonality, this is not included in the presented model and not observed in the model PSD. Thanks to reviewer, who has stimulated us to think how could be empirically observed intra-day pattern of return volatility and trading activity introduced into our model. The answer is not so simple as one could expect because we use simplified, assumed as only statistically valid, relation between log price $p(t)$ and trading activity, see Eq.(\ref{eq:taunfxi}). The proposed model does not include trading activity as independent dynamic variable and this makes introduction of discussed seasonality regarding trading activity not straightforward. Nevertheless, this intraday seasonality first of all has to be reflected  in volatility pattern of total return fluctuations quantified by parameter $b$. Let us to replace constant value of $b=1$ by some intra-day exponential variation
\begin{equation}
b(t) = \exp{[-(t-195)^2/w^2]}+0.5, \label{eq:bseasonality}
\end{equation}
where $w=20$ quantifies in minutes the width of $b$ variation and time $t$ is closed in the circle of total duration of NYSE trading day equal to 390 minutes, note $t=0$ or $t=390$ corresponds to the middle point of trading day. In Fig. \ref{figure6} we present results of such numerical experiment, where power spectral density of absolute return for NYSE stocks is compared with model accounting for $b$ seasonality, Eq. (\ref{eq:bseasonality}). As one can see, such a simple introduction of volatility seasonality reproduces empirical resonance structure of absolute return very well.

For such emerging markets as Baltic region the intensity of stock trading is up to 50 time lower than in NYSE. This makes a considerable impact on return in such short time interval as 1 minute PDF as for over 90\% of time intervals no transactions are registered. Consequently, one observes considerable discrepancy in PDF for low return values as main part of returns are equal to zero. Nevertheless, it is worth to notice that  power law part of PDF and PSD is in pretty good agreement with model one. Mentioned discrepancy disappears with increasing $T$ or for the markets with a more intensive trading. Some PDF discrepancy for low returns values is observed for NYSE and Warsaw stocks as well and this is related with some impact of discrete nature of price values measured in cents, namely on the discrete tick size.  Certainly, prices are smooth in the model.

Finally we can conclude that three state herding model of absolute return in financial markets works very well and explains general origins of power law statistics for very different markets starting from most developed to emerging.

\section{Conclusions}

The proposed consentaneous agent-based and stochastic model of the financial markets is a result of our previous research in stochastic modeling, see references in \cite{Gontis2010Intech,Kononovicius2012IntSys}, and agent-based modeling of herding interaction \cite{Kononovicius2012PhysA,Kononovicius2013EPL}. The idea to combine
two approaches comes from the properties of nonlinear stochastic differential equations generating power law statistics \cite{Gontis2010Intech,Gontis2012ACS} and possibility to derive these equations as macroscopic outcome of microscopic herding model \cite{Kononovicius2012PhysA}. The possibility to reproduce power law statistical features of absolute return observed in the financial markets in such details as power spectrum with two different exponents distinguishes this approach as compromise of sophistication and simplicity.  Nevertheless, the main value of this model comes form its very clear microscopic background of herding interactions between agents.  Very general idea, which has roots in the entomological studies of ant colonies \cite{Kirman1993QJE}, can be adopted to build
system of three agent groups acting in the financial markets. Further complexity of the behavioral aspects of agents can be treated as irrelevant for the macroscopic outcome of this complex system as herding alon reproduces statistical properties.

It is worth to notice that some feedback of observed return increases trading activity in the market and the degree of nonlinearity, which we account by empirical value of the exponent $\alpha=2$, \cite{Gabaix2003Nature,Gabaix2006QJE,Farmer2004QF}. Other model parameters are less grounded by the empirical market analysis and are defined by adjusting proposed model to the here presented statistical properties of the considered financial markets. Parameters $\varepsilon_{cf}=0.1$ and $\varepsilon_{fc}=3$ define the most fundamental tradeoff between fundamental and speculative behavior of agents in financial markets. It is clear from the defined values of the parameters, exhibiting strong asymmetry between fundamentalism and chartism, that bubbles in the financial markets can be explained just by the disappearance of the fundamental trading behavior and markets become stable when considerable part of agents return to the trading according to the fundamental values. The tradeoff between fundamental and speculative behavior is the slowest process in the model defined by adjusted parameter $h=10^{-8} s^{-1}$. It looks like that this long term herding process has global nature and impacts all markets in the similar way.

Much more rapid process defined by parameter $H=300$ describes tradeoff between choices to sell or buy and probably is more related with local stock price dynamics. Nevertheless, the same value of $H$ appropriate for all markets and all stocks makes some surprise and probably is related with universal behavior of power spectral density. Notice that value $\varepsilon_{cc}=3$ is higher than 1 making these rapid fluctuations symmetric and localized around equilibrium $\xi=0$.

From our point of view this study based on the concepts of statistical physics contributes to the behavior finance supporting the general idea of market inefficiency \cite{Akerlof2009Princeton,Shiller2012Princeton} as exhibits possible dominance of herding interactions over agent's rationality.  In other words, rationality is too heterogeneous to resist herding tendencies. This rises the question whether  markets are able to determine true values of assets or one needs other more fundamental sources of economic information. Fortunately the tendencies of imitation open the new possibility that herding itself can be used to stabilize unwanted fluctuations of the financial markets \cite{Kononovicius2014PhysA}.

\section{Acknowledgments}

The authors wish to thank Dr. Rafal Rak, Saulius Malinauskas and Giedrius Bacevicius for kind collaboration.
We thank our colleagues Prof. Bronislovas Kaulakys and Dr. Julius Ruseckas for permanent interest in our work.


\end{document}